\documentclass[12pt,aaspp4]{aastex}

\shortauthors{JONES, ET AL.}
\shorttitle{THE JETS AND DISK IN NGC 4261}

\begin{document}
%
%
\title{The Radio Jets and Accretion Disk in NGC 4261} 

\author{Dayton L.~Jones} 
\affil{Jet Propulsion Laboratory, California Institute of Technology,
Mail Code 238-332, 4800 Oak Grove Drive, Pasadena, CA 91109}
\email{dj@bllac.jpl.nasa.gov} 

\author{Ann E.~Wehrle}
\affil{Jet Propulsion Laboratory, California Institute of Technology, 
Mail Code 301-486, 4800 Oak Grove Drive, Pasadena, CA 91109}
\email{aew@ipac.caltech.edu}

\author{David L.~Meier}
\affil{Jet Propulsion Laboratory, California Institute of Technology,
Mail Code 238-332, 4800 Oak Grove Drive, Pasadena, CA 91109}
\email{dlm@cena.jpl.nasa.gov}

\and

\author{B.~Glenn Piner}
\affil{Jet Propulsion Laboratory, California Institute of Technology,
Mail Code 238-332, 4800 Oak Grove Drive, Pasadena, CA 91109}
\email{glenn@herca.jpl.nasa.gov}

\newpage

\begin{abstract}  
The structure of AGN accretion disks on sub-parsec scales can be probed
through free-free absorption of synchrotron emission from the base of
symmetric radio jets.  For objects in which both jet and counterjet are
detectable with VLBI, the accretion disk will cover part of the counterjet
and produce diminished brightness whose angular size and depth as a function
of frequency can reveal the radial distribution of free electrons in the 
disk.  The nearby (41 Mpc, independent of ${\rm H}_{0}$) FR-I radio galaxy 
NGC 4261 contains a pair of 
symmetric kpc-scale jets.  On parsec scales, radio emission from the nucleus 
is strong enough for detailed imaging with VLBI.  We present new VLBA
observations of NGC 4261 at 22 and 43 GHz, which we combine with previous
observations at 1.6 and 8.4 GHz to map absorption caused by an inner 
accretion disk.
The relative closeness of NGC 4261 combined with the 
high angular resolution provided by the VLBA at 43 GHz
gives us a very high linear resolution,
approximately $2 \times 10^{-2}$ pc $\approx$ 4000 AU $\approx$ 400 
Schwarzchild radii 
for a $5 \times 10^{8} \ {\rm M}_{\odot}$ black hole.  
The jets appear more symmetric at 1.6 GHz
because of the low angular resolution available.  The jets are also more 
symmetric at 22 and 43 GHz, presumably because the optical depth of 
free-free absorption is small at high frequencies.  
At 8.4 GHz neither confusion effect is dominant and absorption 
of counterjet emission by the presumed disk is detectable. 
We find that the orientation 
of the radio jet axis is the same on pc and kpc scales, indicating that the
spin axis of the inner accretion disk and black hole has remained unchanged
for at least $10^{6}$ (and more likely $> 10^{7}$) years.  
This suggests that a single merger event may be responsible for the 
supply of gas in the nucleus of NGC 4261.
The jet opening angle is between $0.3^{\circ}$ and $20^{\circ}$ during
the first 0.2 pc of the jet, and must be $< 5^{\circ}$ during the 
first 0.8 pc.
Assuming that the accretion disk is geometrically and optically thin and 
composed of a uniform $10^{4}$ K plasma, 
the average electron density in the inner 0.1 pc of the disk 
is $10^{3} - 10^{8}$ cm$^{-3}$.
The mass of ionized gas in the inner pc of the disk is $10^{1} - 10^{3}$ 
${\rm M}_{\odot}$, sufficient to power the radio source for $\sim 
10^{4} - 10^{6}$ years.    
Equating thermal gas pressure and magnetic field strength gives a disk
magnetic field of $\sim 10^{-4} - 10^{-2}$ gauss at 0.1 pc.
We include an appendix containing expressions for a simple, optically
thin, gas pressure dominated accretion disk model which may be 
applicable to other galaxies in addition to NGC 4261.
\end{abstract}

\keywords{accretion, accretion disks --- galaxies: active --- 
galaxies: individual (NGC~4261, 3C270) --- galaxies: jets --- 
galaxies: nuclei}

\newpage

\section{Introduction}

The structure of accretion disks in the inner parsecs of active galactic nuclei
can be probed on sub-parsec scales through VLBI observations of free-free
absorption of synchrotron radiation from the base of radio jets.  The nearby
radio galaxy NGC 4261 (3C270) contains a pair of highly symmetric kpc-scale
jets \citep[]{bd85}, and optical imaging with 
HST has revealed a large ($\sim 300$ pc), nearly edge-on nuclear disk of gas 
and dust \citep[]{ffj96}.  This suggests 
that the radio axis is close to the plane
of the sky.  Consequently, relativistic beaming effects should be negligible.
This orientation also precludes gravitational lensing by the central black hole
from affecting the observed jet-to-counterjet brightness \citep[]{bw97}. 
In addition, the central milliarcsecond (mas) scale radio source is strong
enough for imaging with VLBI \citep[]{jst81}. 
The combination of high angular resolution provided by VLBI at 43 GHz
and the relative closeness of NGC 4261 (41 Mpc; \citet[]{f89}) give 
us a very high linear resolution,
approximately 0.1 mas $\approx$ 0.02 pc $\approx$ 4000 AU $\approx$ 400 
Schwarzchild radii for a $5 \times 10^{8} \ {\rm M}_{\odot}$ black hole.  
Thus, NGC 4261 is a good system both for the study of an edge-on accretion disk 
and for intrinsic differences between a jet and counterjet.  

Our previous 8.4 GHz VLBA image of NGC 4261 (Figure 4 in \citet[]{jw97}, 
hereafter JW; reproduced here as Figure 1) showed the inner parsec of the jet 
and counterjet, including a surprising narrow gap in emission at the base of 
the counterjet which we suspected 
was caused by free-free absorption in an accretion disk.  

\placefigure{fig1}

The central brightness
peak at 8.4 GHz was identified as the core by JW based on its inverted spectral
index between 8.4 and 1.6 GHz.  The gap seen at 8.4 GHz is similar to the 
even more dramatic gap seen in the center of NGC 1052 (0238-084) at 15 GHz 
by \citet[]{ke98}.   
We report here 22 and 43 GHz VLBA observations which were made to 
map this gap with higher resolution.  
Our goals were to derive the physical characteristics of the disk, namely
the allowed ranges of thickness, diameter, electron density, and magnetic
field strength.  

\section{Observations}

We observed NGC 4261 using 9 stations of the VLBA with full tracks
on 7 September 1997, 
alternating 22-minute scans between 22.2 and 43.2 GHz.  
Good quality data were obtained from the Hancock, Fort Davis, Pie Town,
Kitt Peak, Owens Valley, and Mauna Kea station, while rain affected 
North Liberty and Los Alamos and technical problems occurred at Brewster. 
The Saint Croix antenna was stowed due to hurricane Erika.  
Left circular polarization was recorded at both frequencies, with a
total bandwidth of 64 MHz.  
Phase offsets between frequency channels were determined and corrected 
using both 3C273 and 1308+326.   

The data were calibrated and fringe-fit using standard routines in 
AIPS\footnote{The Astronomical Image Processing System was developed by
the National Radio Astronomy Observatory.} and imaging, deconvolution, 
and self-calibration were carried out in Difmap \citep[]{spt94}. 
Amplitude calibration of VLBI data at 22 and 43 GHz is often problematic
due to rapidly changing tropospheric water vapor content.  We checked our
{\it a priori} amplitude calibration by comparing short-baseline flux 
densities of 3C273 
with total flux density measurements made at 22 and 37 GHz with the 14-m
Mets\"aehovi antenna.  In both cases our flux densities agree with those
from Metsaehovi to within 15\%.  

An 8.4 GHz image of NGC 4261, made in a similar 
manner from VLBA data obtained in April 1995, is shown in Figure 2 for 
comparison with our newer, higher frequency images.  
The beam size at 8.4 GHz was $1.84 \times 0.80$ mas with the major axis 
almost exactly north-south.
Figure 3 illustrates a possible geometry of the central region which is
consistent with the radio morphology seen in Figures 1 and 2.

\placefigure{fig2}

\placefigure{fig3}

The full resolution (uniform weighting, no taper) VLBA images at 
22 and 43 GHz
are shown in Figures 4 and 5, respectively.  The beam sizes at 22 and
43 GHz are $1.06 \times 0.29$ mas and $0.61 \times 0.16$ mas, with the
major axis within 20$^{\circ}$ of north-south.  We also made naturally
weighted images at both frequencies to search for more extended emission, 
but no detectable emission was found outside of the area shown in 
Figures 4 and 5. 

\placefigure{fig4}

\placefigure{fig5}

In addition, Figure 6 shows the 43 GHz image after convolution with the 
same restoring beam as used in Figure 4 to allow spectral index 
measurements.  The same field of
view is shown in Figures 4 and 6.

\placefigure{fig6}

The wide range of baseline lengths provided by the VLBA should make  
it possible to obtain sufficient overlap of the (u,v) coverage between 
22 and 43 GHz for spectral index determinations.  
However, the correct offset between our 22 and 43 GHz images is not
known {\it a priori} because we do not have absolute positions.  
A range of plausible offsets was tried.  However, we found that 
relatively small changes in offset (less than half the E-W beam width) 
produce large changes in the spectral index map.  Offsets which do 
not produce unreasonably large or small spectral index values tend
to show that the most inverted spectral slope occurs at the position
of the presumed accretion disk absorption and not at the position of
the bright compact core, and that the spectrum becomes steeper away
from the central core and accretion disk region.  However, our data
do not constrain the spectral index distribution in detail.  

To see if amplitude self-calibration had significantly changed the flux 
density scales of our 22 or 43 GHz images, we compared the peak surface 
brightness from images made with only phase self-calibration and images 
made with full amplitude and phase self-calibration.  The resulting amplitude  
corrections applied to the data were 19$\%$ at 22 GHz and 12$\%$ at 43 GHz. 
Uncertainties in the absolute flux density scales can shift all of the 
values in a spectral index map by a constant amount,  
but will not change the shape of the distribution.  

\section{Results}

\subsection{The Parsec-scale Radio Jets}

The position angle of the jets is ${87^{\circ}} \pm {8^{\circ}}$ in our 
VLBI images, and ${88^{\circ}} \pm {2^{\circ}}$ on VLA images \citep[]{bd85}. 
The orientation of the jet axis remains the same on kpc and sub-pc scales,
indicating that the spin axis of the inner accretion disk and black hole 
has remained unchanged for at least $10^6$ years (assuming an average 
expansion speed of 0.1 c), and possibly much longer.

A comparison of our 8, 22, and 43 GHz VLBA images indicates that the 
region just east of the core, including the first half parsec of the 
counterjet, has a highly 
inverted spectrum ($\alpha > 0$, where $S_{\nu} \propto \nu^{\alpha}$).  
It is plausible that free-free absorption by gas in the
central accretion disk is responsible for this.  The jet and counterjet both
have steep radio spectra ($\alpha < 0$) far from the core, as expected.  
Note that if the free-free absorption model is correct, the most inverted 
spectrum may not be located at the position of the true core 
(the ``central engine").

We can set an upper limit for the jet opening angle by noting that
the jet appears unresolved in the transverse (north-south) direction
out to at least 4 mas from the absorption feature in Figure 2.  Using
one quarter of the N-S beam width as an upper limit to the extent of
emission in the transverse jet direction gives an upper limit of 
$5^{\circ}$ for the full opening angle during the first 0.8 pc of
the jet.  A lower limit for the opening angle can be obtained by
requiring that the angular size of emission at the location of the 
bright peak 1 mas from the absorption feature
in Figure 2 be large enough to avoid synchrotron self-absorption. 
This requires an opening angle of $> 0.3^{\circ}$ during the first
0.2 pc of the jet.  Since we believe that the radio jets in NGC 4261
are nearly perpendicular to our line of sight, projection effects 
should be minimal and the intrinsic jet opening angle $\theta$ is 
$0.3^{\circ} < \theta < 20^{\circ}$ during the first 0.2 pc of the
jet and $\theta < 5^{\circ}$ during the first 0.8 pc of the jet.

Since emission from both the jet and counterjet is detectable with VLBI, 
it may be possible to measure proper motions on both sides of the core 
in this source.  If so, the orientation of the radio jets with respect 
to our line of sight can be found, and any 
resulting small relativistic beaming effects on the jet/counterjet
brightness can be taken into account.  A more sensitive 43 GHz VLBI image 
(where free-free absorption is minimal) can then be used to see just 
how similar the jet acceleration and collimation processes are on both 
sides of a ``central engine" at the same epoch.

\subsection{The Inner Accretion Disk}

The fact that our 1.6 GHz image (see Figure 1 in JW) is highly symmetric 
lets us set an upper limit to the angular size of any absorption feature.
At this frequency the free-free optical depth should be large, so to  
avoid detection the angular size of the absorption feature must be much 
smaller than our angular resolution at 1.6 GHz.  The restoring beam in 
Figure 1 of JW has an east-west size of 9 mas.  The jet/counterjet 
brightness ratio is unity from the core out to at least $\pm 25$ mas.  
Using the observed brightness ratio of $\approx 1$ at
$\pm 10$ mas ($\sim \pm 2$ pc), just larger than our resolution, tells
us that the transverse size of any deep absorption feature is $<< 2$ pc.  
It is expected that the inner pc or so of an accretion disk orbiting a 
massive ($\sim 10^{8}-10^{9} \ {\rm M}_{\odot}$) black hole will be
geometrically thin, and consequently we will assume a typical disk
thickness of $<< 0.1$ pc and a nominal line-of-sight path length through
the inner disk of $\sim 0.1$ pc. 
We now use the HI and CO column densities measured by \citet[]{jm94},  
$\approx 10^{21} \ {\rm cm}^{-2}$, to estimate an 
electron number density of 
$n_{e} \ge 3 \times 10^{3} \ (0.1/L) \ {\rm cm}^{-3}$, where $L$ is the 
path length in pc.  A slightly lower density would be deduced using the 
X-ray absorption column density 
of ${\rm N}_{\rm H} < 4 \times 10^{20}$ cm$^{-2}$ from 
\citet[]{wb94}.  In both cases it is 
assumed that there will be on average one free electron per nucleus.  
This is a conservative assumption since the inner 0.1 pc of the disk 
should be highly ionized and 
have a density at least as large as the outer neutral regions. 
The inclination of the disk is needed to further constrain the path 
length $L$ and thus the average electron density $n_{e}$.

To get an upper limit for $n_{e}$, we note that the jet/counterjet 
brightness ratio at 43 GHz near the core in Figure 6 is also small 
($\approx 2$).  This implies
a low optical depth at 43 GHz.  Assuming an electron temperature 
of $\sim 10^{4}$ K 
(plausible at a disk radius of 0.1 pc),  
an optical depth $\tau < 1$ at 43 GHz requires $n_{e} < 4 \times 10^{5} \
\sqrt{0.1/L} \ {\rm cm}^{-3}$.  
However, the temperature of the disk increases at smaller radii, 
possibly reaching $10^{7} - 10^{8}$ K near the inner edge.  At these 
temperatures, an electron number density of $\sim 10^{8} \ 
\sqrt{0.1/L} \ {\rm cm}^{-3}$ is needed to 
produce an optical depth of unity at 43 GHz.  
The upper limit for electron number density will be somewhere 
between $4 \times 10^{5} \ \sqrt{0.1/L}$
and $10^{8} \ \sqrt{0.1/L} \ {\rm cm}^{-3}$, depending on the actual 
range of electron temperatures along the line of sight.

Thus, the electron number density of the inner 
accretion disk, averaged over the line of sight, is
$$3 \times 10^{3} \ (0.1/L) \ < \ n_{e} \ < \ 10^{8} \ \sqrt{0.1/L}
\ \ \ {\rm cm}^{-3}.$$
For comparison, the inner accretion disk believed to be responsible 
for the unusually inverted spectrum of the radio core in Centaurus A 
has an average electron number density of at least $10^{5} \ 
{\rm cm}^{-3}$ for a path length of 1 pc \citep[]{j96}. 
Assuming a thin inner disk in NGC 4261 with a radius of 1 pc and a 
thickness of 0.01 pc, an average electron density of $10^{6}$ cm$^{-3}$, 
and one proton per electron,
the mass of ionized gas in the disk is $\sim 10^{3} \ {\rm M}_{\odot}$.
Of course, a larger assumed thickness or radius would lead to a larger 
inner accretion disk mass.
Even $10^{3} \ {\rm M}_{\odot}$ is sufficient to fuel the central 
engine for
$10^6$ years at an accretion rate of $10^{-3} \ {\rm M}_{\odot}$ 
per year, consistent with the observed luminosity.  
A more realistic disk model with thickness increasing with radius 
and density decreasing with radius is described in Appendix A.
This model predicts a lower total mass of ionized gas within 1 pc,
which implies that material must migrate through the disk
from radii $> 1$ pc during the source lifetime.  It is interesting 
to note that the electron number density we derive at 0.1 pc is 
comparable to that in the solar corona at  
$\sim 2 \ {\rm R}_{\odot}$.  
That is, the disk is tenuous and optically thin to visible light.  

Equating the thermal gas pressure in
the disk with the local magnetic field ($B^{2} = {8\pi} \ \alpha \ n_{e}kT$, 
where $\alpha \approx 0.01$ is the usual viscosity parameter; 
\citet[]{ss73}) gives a disk magnetic field of $\sim
10^{-2} - 10^{-4}$ gauss at 0.1 pc. 
General expressions for estimating physical parameters in an optically thin,
gas pressure dominated accretion disk at radii of $\sim 0.1 - 1$ pc are 
derived in Appendix A.  Table 1 compares these parameter values with those 
for other galaxies in which there is evidence for an inner accretion disk. 
Of the five nearby ($<100$ Mpc) radio galaxies for which some inner accretion 
disk characteristics are known, only two (NGC 4261 and NGC 4258) 
have well-determined central black hole masses (see Table 1).  

\section{Discussion}

A geometrically thin inner disk which would not completely obscure optical 
emission from the core is a plausible model given the very high percentage
of low luminosity 3CR radio galaxies which have bright, unresolved optical 
continuum sources visible on HST/WFPC2 images \citep[]{ccc98}. 
This includes NGC 4261.  A geometrically  
and optically thick dusty torus would obscure the central optical continuum 
source in many low luminosity objects with FR I radio morphology, since
these sources are expected to be oriented at large angles to our line of 
sight.  A nearly edge-on disk orientation may also explain the low bolometric 
luminosity of the NGC 4261 nucleus and lack of an ultraviolet
excess in its spectral energy distribution \citep[]{h99}. 
Additional constraints on the inclination and density of the inner disk 
could come from observations of free-free radio emission from ionized disk 
gas or radio emission from the central jets scattered by electrons in the 
far inner edge of the accretion disk \citep[]{gbo97}. 
However, the dynamic range of our VLBA images is not
adequate to detect this emission in the presence of the bright parsec-scale 
radio jets.  Another way to constrain the radio axis, and thus the inner 
disk, orientation would be measurement of 
proper motions in both the jet and counterjet (e.g., \citet[]{twv98}). 
VLBA observations to attempt this are underway. 

The above analysis makes use of the fact that the observed jet/counterjet 
brightness ratio in NGC 4261, and in some other galaxies which are 
well-observed with VLBI, peaks at intermediate frequencies and falls to 
nearly unity at both low frequencies (where the beam is much larger than 
the angular size of the absorbing material) 
and high frequencies (where the free-free optical depth becomes very small). 
See, for example, Figure 15 in \citet[]{k98}. 
At low frequencies the brightness ratio $R$ should decrease approximately
linearly with frequency ($\theta_{\rm beam} \propto \nu^{-1}$), while at 
high frequencies the fall-off of $R$ should be more rapid 
($\tau_{\rm f-f} \propto 
\nu^{-2}$).  With VLBA images at four frequencies we can not confirm this
behaviour in detail, but it is clear that near the core the brightness 
ratio $R$ is greater at 
8.4 GHz than at 1.6, 22, or 43 GHz.  

The mass of the central black hole in NGC 4261 
is $(5\pm1) \times 10^{8} \ {\rm M}_{\odot}$ \citep[]{ffj96}. 
Thus, the mass 
of material in the inner accretion disk is negligible compared to 
the black hole mass, and the orbital period of material in the 
inner accretion disk at a radius
$r$ (in pc) is 
$\approx 4 \times 10^{9} \ (r/0.1)^{3/2}$ seconds ($\approx 10^{2}$ 
years for $r$ = 0.1 pc).  The spin rate  
of the central black hole is unknown, but is predicted to be small
to moderate by the model of \citet[]{m99}. 
A lower limit on the spin of the black hole can be derived from the 
known mass of the hole, the assumed accretion rate 
($10^{-3} \ {\rm M}_{\odot}$ per year),
and the spin axis alignment timescale ($> 10^{6}$ years) implied by 
the co-linear kpc scale jets.  The resulting dimensionless spin 
parameter (see \citet[]{mtw73}) 
is $> 2 \times 10^{-4}$,
which allows either high or low black hole spin rates. 

The angular momentum of gas in the inner accretion disk is expected 
to be aligned 
with the spin axis of the black hole \citep[]{bp75}. 
If the black hole is spinning slowly, its spin axis will eventually 
become aligned with the 
angular momentum of the accreting gas at large radii 
\citep[]{np98}.  However, the long-term directional 
stability of the radio jets in NGC 4261 implies that the gas falling 
into the central region
of the galaxy and supplying the central engine has had a constant 
angular momentum direction for most of the life of the radio source.  
This in turn implies that a single merger event may be responsible 
for the supply of gas in the nucleus of NGC 4261.

\section{Conclusions}

We have imaged the nuclear radio source in NGC 4261 with the VLBA at 
four frequencies.  We find that the jet/counterjet brightness ratio 
near the core is larger at 8.4 GHz than at lower frequencies. 
This can be explained by a combination of two effects:  low angular  
resolution at 1.6 GHz which masks small-scale brightness variations, 
and low absorption by ionized thermal gas at 22 and 43 GHz.  The 
brightness asymmetry at 8.4 GHz could be caused by a nearly edge-on 
inner accretion disk.  If so, the (model dependent) electron density 
in the inner 0.1 pc of the disk has an average value between $3 
\times 10^{3}$ and 
$10^{8} \ {\rm cm}^{-3}$.  Future observations to measure proper
motions in the jet and counterjet will better define the orientation 
of the radio axis and thus the inclination of the inner accretion 
disk.  This will determine the path length $L$ through the disk, 
and consequently reduce the allowable range of electron number
density averaged over the path length.  

The optically thin, uniform temperature model disk described in 
the appendix, derived only from accretion 
physics and values of the black hole mass and accretion rate consistent 
with NGC 4261, is remarkably similar to the disk that we observe in 
free-free absorption against the radio jet.  We therefore believe  
that these observations have detected the sub-pc 
accretion disk powering the active nucleus in NGC 4261.

\acknowledgements
The Very Long Baseline Array is part of the National Radio
Astronomy Observatory, which is a facility of the National
Science Foundation operated by Associated Universities, Inc.,
under a cooperative agreement with the NSF.  
A.W.~gratefully acknowledges support from the NASA Long Term 
Space Astrophysics Program.
We thank H.~Ter\"asranta for making the Mets\"aehovi flux density
measurements of 3C273 available, and the anonymous referee for 
helpful suggestions.  This research was carried out at the Jet 
Propulsion Laboratory, California Institute of Technology, under 
contract with the National Aeronautics and Space Administration.

\appendix

\label{msd_appendix}

\section{Isothermal, Optically-thin Accretion Disks far from Black Holes}

In this appendix we describe a simple accretion disk model, suitable for 
optically thin gas-pressure dominated flow far from the black hole, in 
which the plasma temperature is a very slowly varying function of distance 
from the hole.
The canonical temperature is assumed to be $10^4 {\rm K}$ for the range of 
interest in this model.  (Such a temperature is typical of an optically thin, 
warm plasma undergoing moderate heating to keep it mostly ionized.)
Following \citet[]{ss73}, and ignoring 
the factor of $(1-r^{-1/2}) \approx 1$ (since at $0.1-1.0$ {\rm pc}, 
$r=R/3R_{Sch} = 3400-34,000$ for a $10^8 {\rm M_{\sun}}$ black hole), 
we obtain the following equations for the disk structure, expressed in 
units of $M_{8} \equiv M_{BH}/10^8 {\rm M_{\sun}}$ for the black hole mass,  
$\dot{M}_{-3} \equiv \dot{M}/(10^{-3} {\rm M_{\sun} \, yr^{-1}}) = 
\dot{M}/(6.3 \times 10^{22} {\rm g \, s^{-1}})$ for the accretion rate, 
and $R_{18} \equiv R/10^{18} {\rm cm}$ for the distance from the black hole.

From the equation for hydrostatic equilibrium in the direction perpendicular 
to the disk, the disk half-thickness is
\begin{equation}
H \; = \; 0.0026 \, {\rm pc} \; M_{8}^{-1/2} \; R_{18}^{3/2}
\end{equation}
For NGC 4261, with $M_{8} = 5$, the {\em full} disk thickness at a radius
of 1 pc ($R_{18} = 3$) is 0.012 pc, very close to the 0.01 pc thickness 
assumed for the simple uniform thickness disk in section 3.2. 
Adopting a radiative cooling rate of $\sim 3 \times 10^{-23} \; n^2 \; 
{\rm erg \, cm^{-3}}$ at $10^4 {\rm K}$ \citep[]{kr99} and 
setting it equal to 
the energy generated locally by viscous (accretion) heating, we 
obtain the particle density
\begin{equation}
n \; = \; 1.5 \times 10^{4} \, {\rm cm^{-3}} \; M_{8}^{3/4} \; 
\dot{M}_{-3}^{1/2} \; R_{18}^{-9/4}
\end{equation}
The approximate optical depth of the disk in a direction perpendicular 
to it, due to the inverse absorption processes
(at $\nu \sim kT/h \approx $ optical frequencies), is 
\begin{equation}
\tau_{abs} \; = \; 1.4 \times 10^{-11} \; M_{8} \; 
\dot{M}_{-3} \; R_{18}^{-3}
\end{equation}
and that due to electron scattering is
\begin{equation}
\tau_{es} \; \leq \; 1.5 \times 10^{-4} \; M_{8}^{1/4} \; 
\dot{M}_{-3}^{1/2} \; R_{18}^{-3/4}
\end{equation}
(an upper limit, since the gas is not fully ionized).  
At IR-optical-UV wavelengths, therefore, the disk is very optically 
thin.  It becomes optically thick to free-free absorption in the 
perpendicular direction ($\kappa_{\nu} \, m_p \, n \, 2 H \geq 1$) 
only below the frequency
\begin{equation}
\nu_{perp} \; = \; 250 \, {\rm MHz} \; M_{8}^{1/2} \; 
\dot{M}_{-3}^{1/2} \; R_{18}^{-3/2}
\end{equation}
However, if viewed nearly edge-on, with the line-of-sight passing 
roughly from $R_{18}/3$ to
$3 R_{18}$ ({\it i.e.}, $0.1-1.0 \, {\rm pc}$), the disk will be 
optically thick to free-free 
($\int \kappa_{\nu} \, m_p \, n \, dR \geq 1$) below the frequency
\begin{equation}
\nu_{edge-on} \; = \; 8 \, {\rm GHz} \; M_{8}^{3/4} \; 
\dot{M}_{-3}^{1/2} \; R_{18}^{-7/4}
\end{equation}
For $M_{8} = 5$ we get 
\begin{equation}
\nu_{edge-on} \; = \; 27 \, {\rm GHz} \; 
\dot{M}_{-3}^{1/2} \; R_{18}^{-7/4}
\end{equation}
The mass of the disk within this radius range, and flow time scale 
($R/v_{R}$), are
\begin{eqnarray}
M_{disk} \; = \; 3.9 \, {\rm M_{\sun}} \; M_{8}^{1/4} \; 
\dot{M}_{-3}^{1/2} \; R_{18}^{5/4}
\\
t_{acc} \; = \; 1200 \, {\rm yr} \; M_{8}^{1/4} \; 
\dot{M}_{-3}^{-1/2} \; R_{18}^{5/4}
\end{eqnarray}

Three consistency checks on the model are the ratio of the 
half-thickness to disk radius at a given point, the ratio of the  
accretion (inflow) velocity to the Keplerian velocity, 
and the ratio of inward heat advection to radiative cooling, all 
of which should be small, given the assumptions of the model.  
We find that
\begin{equation}
\frac{H}{R} \; = \; 0.008 \; M_{8}^{-1/2} \; R_{18}^{1/2}
\end{equation}
for the disk height, 
\begin{equation}
\frac{-v_{R}}{v_{K}} \; = 0.23 \; M_{8}^{-3/4} \; 
\dot{M}_{-3}^{1/2} \; R_{18}^{1/4}
\end{equation}
for the inflow velocity, and 
\begin{equation}
\frac{\dot{\varepsilon}_{adv}}{\dot{\varepsilon}_{rad}} \; = \; 
2.5 \times 10^{-4} \; M_{8}^{-1} \; R_{18}
\end{equation}
for the magnitude of the advection of thermal energy.  
Each of these three ratios will be even smaller if $M_{8} = 5$, the
value appropriate for NGC 4261, is used. 
(Note that, because of the low optical depth, the photon 
distribution is not a black body and has an even lower energy 
density than the thermal gas.)  The assumption of a thin, radiating,  
$10^4 {\rm K}$, Keplerian accretion disk, therefore, is consistent  
with the model calculations, as well as with our observational
results for the absorbing disk in NGC 4261. 


\newpage

\figcaption{Grey scale image of NGC~4261 at 8.387 GHz showing 
the bright core and first parsec of the main jet (pointing west) 
plus fainter emission farther along the main jet and near the 
base of the (east pointing) counterjet.  Note the narrow region of
reduced brightness just to the east of the bright core, at the
apparent base of the counterjet.  From \citet[]{jw97}.   
\label{fig1}}

\null

\figcaption{VLBA image of NGC~4261 at 8.387 GHz.
The contour levels are 1, 2, 4, 8, 16, 32, 
and 64\% of the peak surface brightness (101 mJy/beam).  The restoring
beam is $1.84 \times 0.80$ mas with the major axis along position angle
-1.1$^{\circ}$.
\label{fig2}}

\null

\figcaption{Schematic illustration of the central few parsecs in
NGC 4261 showing the relative positions of the core (black hole),
jet, counterjet, and inner accretion disk.
\label{fig3}}

\null

\figcaption{Full resolution 22-GHz VLBA image, showing the first parsec of 
the radio jet and counterjet as well as the bright, barely resolved core.  
The jet-to-counterjet brightness ratio at $\pm 1$ mas from the core is
significantly smaller at 22 GHz than at 8.4 GHz (see Figures 1 and 2). 
The contours are -1, 1, 2, 4, 8, 16, 32, and 64\% of
the peak (165 mJy/beam), and the restoring beam is $1.06 \times 0.29$
mas with the major axis along position angle -18.3$^{\circ}$.
The total flux density in the image is 388 mJy.
\label{fig4}}

\null

\figcaption{Full resolution 43-GHz VLBA image.  The contours are -1,
1, 2, 4, 8, 16, 32, and 64\% of the peak (141 mJy/beam), and the 
restoring beam is $0.61 \times 0.16$ mas with the major axis in
position angle -19.5$^{\circ}$.  Figures 2 and 5 have the same
field of view.
The total flux density in the image is 478 mJy.
\label{fig5}}

\null

\figcaption{43-GHz image convolved with the same restoring beam as the
22-GHz image in Figure 4 to allow spectral indices to be measured. 
The contours levels are 1, 2, 4, 8, 16, 32, and 64\% of the peak
(305 mJy/beam) with the major axis along position angle -18.3$^{\circ}$.
Residuals have not been included in this image.  
Note that Figures 4 and 6 cover the same field of view.  
Although only the innermost parts of the jet and counterjet are 
detectable at 43 GHz, the brightness ratio near the core is closer
to unity than it is at lower frequencies.
\label{fig6}}

\newpage

\begin{center}
\begin{tabular}{llllllllll} 
\hline
\multicolumn{10}{c}{Table 1:  Characteristics of Accretion Disks}
   \\ \hline\hline
\multicolumn{1}{c}{Source}
& \multicolumn{1}{c}{Dist.}
& \multicolumn{1}{c}{$S_{\rm jet}$} 
& \multicolumn{1}{c}{Height}
& \multicolumn{1}{c}{Radius}
& \multicolumn{1}{c}{${\rm n}_{e}$} 
& \multicolumn{1}{c}{Disk mag.}
& \multicolumn{1}{c}{${\rm M}_{\rm BH}$}
& \multicolumn{1}{c}{Assumed} 
& \multicolumn{1}{c}{ } \\ 
\multicolumn{1}{c}{name}
& \multicolumn{1}{c}{Mpc$^{\dagger}$}
& \multicolumn{1}{c}{(Jy)$^{\ddagger}$}
& \multicolumn{1}{c}{(pc)} 
& \multicolumn{1}{c}{(pc)}
& \multicolumn{1}{c}{(${\rm cm}^{-3}$)} 
& \multicolumn{1}{c}{field (G)}
& \multicolumn{1}{c}{(${\rm M}_{\odot}$)}
& \multicolumn{1}{c}{tmp.~(K)}
& \multicolumn{1}{c}{Ref.}
\\ \hline
NGC 5128 & 3.5 & 7.0 & $< 1$ & $> 1$ & $> 10^{5}$ & $> 2 \times 10^{-3}$   
& -- & $10^{4}$ & 1 \\ 
NGC 4258 & 6.4 & 0.003 & $< 0.1$ & $\sim 0.3$ & -- & $\le 3 \times 10^{-1}$ & 
$3.5 \times 10^{7}$ & N/A & 2 \\ 
NGC 1052 & 17 & 1.0 & $\sim 0.1$ & $>1$ & $\sim 10^{5}$ & $3 \times 10^{-3}$ 
& -- & $10^{4}$ & 3 \\ 
NGC 4261 & 41 & 0.5 & $\sim 0.01$ & $\sim 0.1$ & $\sim 10^{6}$ & $10^{-4} - 10^{-2}$ & 
$5 \times 10^{8}$ & $10^{4}$ & 4 \\ 
Mkn 348 & 56 & 0.5 v & $\sim 0.1$ & $\sim 1$ & $10^{5}-10^{7}$ 
& $\approx 6 \times 10^{-4}$ & -- & $8 \times 10^{3}$ & 5 \\ 
Mkn 231 & 168 & 0.03 v & $\sim 0.1$ & $\sim 1$ & $10^{5}-10^{7}$ 
& $\approx 6 \times 10^{-4}$ & -- & $8 \times 10^{3}$ & 5 \\ 
Cygnus A & 225 & 1. v & $< 0.3$ & $< 15$ & $> 7 \times 10^{3}$ 
& $> 4 \times 10^{-4}$ & -- & $> 3 \times 10^{3}$ & 6\\ 
1946+708 & 400 & 0.3 & $< 20$ & $< 50$ & $> 2 \times 10^{2}$ 
& $> 10^{-5}$ & -- & $8 \times 10^{3}$ & 7 \\ 
NGC 1275 & $ 700 $ & 30. v & $\sim 0.1$ & $> 25.0$ & $\sim 7 \times 10^{4}$ 
& $8 \times 10^{-4}$ & $\sim 10^{8}$ & $10^{4}$ & 8 \\ 
%
%
\hline\hline
\end{tabular}
\end{center}

\medskip
\noindent{ $^{\dagger}$ \ Using H$_{0}$ = 75 km s$^{-1}$ Mpc$^{-1}$ for
distances given as redshifts.  The four nearest galaxies have published
distances which do not depend on H$_{0}$.}

\noindent{ $^{\ddagger}$ \ ``v" indicates variability} 

\medskip
\medskip
\medskip
\baselineskip 14pt
\noindent Table 1 references: \\ 

\noindent 1. Jones, et al. 1996; Tingay, et al. 1998. \\
2. Herrnstein, et al. 1997; 1998. \\
3. Braatz, et al. 1999; Kellermann, et al. 1999. \\ 
4. This paper. \\
5. Ulvestad, et al., 1999. \\
6. Krichbaum, et al. 1998. \\
7. Peck, et al. 1999. \\
8. Dhawan, et al. 1998. \\ 

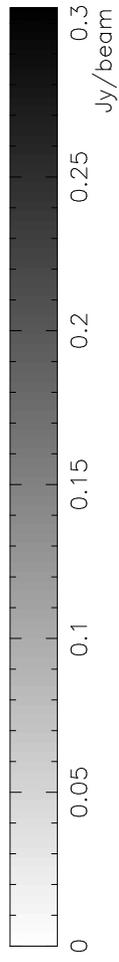
\end{document}